%Paper: hep-ph/9306319
%From: JFGUCD@UCDHEP.UCDAVIS.EDU
%Date: Fri, 25 Jun 1993 13:58 PDT

\input phyzzx.tex
\input tables.tex

\def\hl{h^0}
\def\hh{H^0}
\def\ha{A^0}
\def\mhl{m_{\hl}}
\def\mhh{m_{\hh}}
\def\mha{m_{\ha}}
\def\hpm{H^{\pm}}

\def\hp{H^+}
\def\gam{\gamma}
\def\mstop{M_{\wtilde t}}

\def\gev{~{\rm GeV}}
\def\tev{~{\rm TeV}}
\def\pbi{~{\rm pb}^{-1}}
\def\fbi{~{\rm fb}^{-1}}

%%Journal definitions

\def\prdj#1{{\it Phys. Rev.} {\bf D{#1}}}
\def\npbj#1{{\it Nucl. Phys.} {\bf B{#1}}}

\def\mt{m_t}

\def\wpm{W^{\pm}}
\def\tauptaum{\tau^+\tau^-}
\def\rta{\rightarrow}
\def\tanb{\tan\beta}

\def\hsm{\phi^0}

\def\mhsm{m_{\hsm}}

\def\mz{m_Z}
\def\anti{\overline}

\def\ifmath#1{\relax\ifmmode #1\else $#1$\fi}

\def\3quarter{{\textstyle{3 \over 4}}}

\def\h{h}
\def\mh{m_{\h}}

\input phyzzx
\Pubnum={$\caps UCD-93-20$\cr $\caps SMU-HEP-93/09$ \cr
$\caps SLAC-PUB-6277$\cr $\caps RU-93-23$\cr}
\date{June, 1993}

\titlepage
\vskip 0.75in
\baselineskip 0pt
\hsize=6.5in
\vsize=8.5in
\centerline{{\bf Guaranteed Detection
of a Minimal Supersymmetric Model}}
\centerline{{\bf Higgs Boson at Hadron Supercolliders}}
\vskip .075in
\centerline{J. Dai$^a$, J.F. Gunion$^b$ and R. Vega$^c$}
\vskip .075in
\centerline{\it a) Dept. of Physics and Astronomy, Rutgers University,
Piscataway, NJ 08855}
\centerline{\it b) Davis Institute for High Energy Physics,
Dept. of Physics, U.C. Davis, Davis, CA 95616}
\centerline{\it c) Dept. of Physics, Southern Methodist University,
Dallas, TX 75275}
\centerline{\it and}
\centerline{\it Stanford Linear Accelerator Center, Stanford, CA 94305}

\vskip .075in
\centerline{\bf Abstract}
\vskip .075in
\centerline{\Tenpoint\baselineskip=12pt%\parindent=1pc
\vbox{\hsize=12.4cm
\noindent We demonstrate that expected efficiencies and purities
for $b$-tagging at SSC/LHC detectors should allow detection of
at least one of the Higgs bosons of the Minimal Supersymmetric Model
in $t\anti t$~Higgs production, with Higgs$\rta b\anti b$
decay, over a substantial range of supersymmetric parameter space.
In particular, with the addition of this mode to those previously
considered, there is no region of supersymmetric parameter space
for which {\it none} of the Higgs bosons of the model
can be seen at the SSC/LHC.
}}

\vskip .15in
\noindent{\bf 1. Introduction and Procedure}
\vskip .075in

Understanding the Higgs sector is a crucial
mission for future high energy colliders such
as the SSC and LHC. While it is quite certain that the SSC/LHC
will be able to detect the Standard Model Higgs boson ($\hsm$),
prospects for detection of the Higgs bosons of the Minimal Supersymmetric
Model (MSSM) have appeared to be somewhat limited.  In particular,
\REF\everybody{J.F. Gunion and L.H. Orr, \prdj{46} (1992) 2052.
Z. Kunszt and F. Zwirner, \npbj{385} (1992) 3.
H. Baer, M. Bisset, C. Kao and X. Tata,
\prdj{46} (1992) 1067.
V. Barger, K. Cheung, R.J.N. Phillips and A.L. Stange,
\prdj{46} (1992) 4914.}
\REF\pergunion{For a review see J.F. Gunion, preprint UCD-92-20, to appear in
{\it Perspectives in Higgs Physics}, ed. G. Kane,
World Scientific Publishing (1992).}
in previous work\refmark{\everybody,\pergunion}
it became apparent that for $\mt\sim150\gev$ there is a window
of $\mha$--$\tanb$ parameter space, with $110\lsim\mha\lsim 170\gev$
and $\tanb\gsim 10$, in which it could be that
no MSSM Higgs boson would be seen either at LEP-II or at the SSC/LHC.
(There is an even larger window where none could be seen
just at the SSC/LHC.)
For the former window to arise it is necessary that the squark mass be
large enough ($\sim 1\tev$) that radiative corrections
\Ref\perhaber{For a review, see H.E. Haber, preprint UCSC-92/31, to appear in
{\it Perspectives in Higgs Physics}, ed. G. Kane,
World Scientific Publishing (1992).}
to the mass of the lightest CP-even Higgs boson ($\hl$) are big, in particular,
big enough that $Z^*\rta Z\hl$ is not kinematically allowed at LEP-II
when $\tanb$ is large and $\mha$ takes on moderate to large values.
In addition, the presence of this window of unobservability
is based on the assumption that only the
$\gam\gam$ and $4\ell$ decay channels of the
CP-even $\hl$ and $\hh$ Higgs bosons will lead to viable
signals at the SSC/LHC,
\foot{Detection of the CP-odd $\ha$ in the $\gam\gam$ or $4\ell$
decay channels is confined to the region of parameter space where
$\tanb\lsim 1$.}
and that detection of the $\hpm$ will be possible
only if $t\rta\hp b$ decays are kinematically allowed.
Although it is not impossible that the portion of the window
with very high $\tanb$ (where inclusive $b\anti b$ fusion yields enhanced
Higgs boson cross sections) can be covered using the $\tauptaum$ decay
modes of the $\hh$ and the $\ha$,
\Ref\kz{Z. Kunszt and F. Zwirner, Ref.~\everybody.}
the precise parameter
region over which it can be employed is a subject of continuing study.
\Ref\lhcstudies{See the LHC detector design studies.}
In this letter, we demonstrate that a viable signal at the SSC/LHC for
the primary decays $\hl\rta b\anti b$ or $\hh\rta b\anti b$ can be obtained
in $t\anti t$~Higgs production over much of $\mha$--$\tanb$
parameter space (in particular, in the window in question).

In a previous paper,
\Ref\previous{J. Dai, J.F. Gunion, and R. Vega, preprint UCD-93-18
(1993).}\ we established that detection of the SM $\hsm$
in the $t\anti t b\anti b$ final state
mode was, indeed, possible for a significant range of $\mhsm$,
provided $\mt\gsim 130-140\gev$.
The analysis of the present paper is based on the results of this previous
work, to which we refer the reader for details. Several different
detection scenarios were examined there.  Here we consider only
the case in which 3 $b$-quarks are required to be vertex tagged
with efficiency of 30\% and purity of 1\% (after appropriate
kinematic and vertex separation cuts).  For this scenario, termed
case I in Ref.~\previous, substantial peaks in the $b\anti b$
mass spectrum are visible for $\hsm$ masses in the region below
about $110-120\gev$. Of course, as argued
in Ref.~\previous, it might be that the above assumptions regarding
$b$-tagging efficiency and purity are too conservative.
Detection of the MSSM Higgs bosons in the $t\anti t b\anti b$ final state
mode would be even easier if, for instance, the $b$-tagging probability could
be brought up to 40\% while decreasing the misidentification
probability to 0.5\% over the stated kinematic range.

To convert the case I results for use in the MSSM, the following procedure
is employed.  First, we give the statistical significance
of the $b\anti b$ mass peaks at Higgs masses of 80, 100, 120 and 140 GeV,
for $\mt=110$, 140 and 180 GeV, for $L=10\fbi$  at the SSC
{\it assuming 100\% $b\anti b$ branching ratio}.\foot{Results for the LHC
will not be given in this letter, but are substantially similar to
those obtained for the SSC.}
The number of standard deviations obtained for $\mt=(110,140,180)$
is (4.0,5.0,10.2), (3.2,3.4,6.5), (2.7,3.1,3.9) and (2.1,2.4,3.1)
at Higgs masses of 80, 100, 120 and 140 GeV, respectively.
Interpolation/extrapolation is employed to obtain results for
the top quark masses of 150 GeV and 200 GeV considered here,
and for other values of Higgs mass.  The appropriate
result for the $\hl$ or $\hh$ is obtained by multiplying
the number of standard deviations
by the $b\anti b$ branching ratio of the $\hl$ or $\hh$ and
by the square of the ratio of the $\hl$ or $\hh$
$t\anti t$ coupling to the $\hsm t\anti t$ coupling
(to account for the difference in the $t\anti t$~Higgs production rates).
Finally, if $t\rta\hpm b$ decays are allowed, the rate for the $t\anti t$~Higgs
signal and for all the $t\anti t$-related backgrounds (see Ref.~\previous)
must be appropriately reduced to account for the reduced trigger rate
of an isolated lepton from one or two $t\rta \wpm b$ decays.
Clearly, all these correction factors, not to mention the Higgs masses
themselves, depend upon the supersymmetric model parameters, and in particular
on $\mha$ and $\tanb$.

\FIG\couplings{The ratios $g_{t\anti t \h}^2/g_{t\anti t\hsm}^2$ and
$g_{b\anti b \h}^2/g_{b\anti b\hsm}^2$ ($\h=\hl,\hh$) are plotted
as a function of $\mha$ at $\tanb=2$ and $\mt=150\gev$.
We have taken $\mstop=1\tev$.
}

\midinsert
\vbox{\phantom{0}\vskip 4.5in
\phantom{0}
\vskip .5in
\hskip +30pt
\special{ insert user$1:[jfgucd.rcsusyhiggs]hbb_susy_fig1.ps}
\vskip -1.65in }
{\rightskip=3pc
 \leftskip=3pc
 \Tenpoint\baselineskip=12pt%\parindent=1pc
\noindent Figure~\couplings:
The ratios $g_{t\anti t \h}^2/g_{t\anti t\hsm}^2$ and
$g_{b\anti b \h}^2/g_{b\anti b\hsm}^2$ ($\h=\hl,\hh$) are plotted
as a function of $\mha$ at $\tanb=2$ and $\mt=150\gev$.
We have taken $\mstop=1\tev$.
}

\endinsert
%%%%

\FIG\brbbbar{$BR(\h\rta b\anti b)$ ($\h=\hl$ or $\hh$)
as a function of $\mh$ at $\mt=150\gev$, for $\tanb=2$ and 5.
We have taken $\mstop=1\tev$. Decays to chargino and neutralino
pair states are assumed to be forbidden.
}

\midinsert
\vbox{\phantom{0}\vskip 4.5in
\phantom{0}
\vskip .5in
\hskip +30pt
\special{ insert user$1:[jfgucd.rcsusyhiggs]hbb_susy_fig2.ps}
\vskip -1.65in }
{\rightskip=3pc
 \leftskip=3pc
 \Tenpoint\baselineskip=12pt%\parindent=1pc
\noindent Figure~\brbbbar:
$BR(\h\rta b\anti b)$ ($\h=\hl$ or $\hh$)
as a function of $\mh$ at $\mt=150\gev$, for $\tanb=2$ and 5.
We have taken $\mstop=1\tev$. Decays to chargino and neutralino
pair states are assumed to be forbidden.
}

\endinsert
%%%%

To understand the results we shall obtain, it is useful to review some features
of the MSSM Higgs sector.
\Ref\hhg{J.F. Gunion, H.E. Haber, G. Kane and S. Dawson,
{\it The Higgs Hunter's Guide}, Addison-Wesley, Redwood City, CA (1990).}
First, recall that the $\hl$ and $\hh$ mass
eigenstates are obtained by diagonalizing the CP-even  mass-squared matrix.
The result for $\tanb\gsim 2$ is a kind of level-crossing in which
for low $\mha$ the $\hh$ has a relatively constant mass
value somewhat above $\mz$.
As $\mha$ increases, $\mhl$ rises to `meet' this constant value.
For $\mha$ values above the crossing point, $\mhl$ takes on
a value which is a bit below or nearly the same (at large $\tanb$)
as the constant value while $\mhh$ rises, becoming approximately
degenerate with $\mha$. For $\mt\sim 150\gev$ and $\mstop\sim 1\tev$,
the constant mass value referred to above is in the vicinity of 100 GeV,
\foot{The precise number depends upon $\tanb$.}
\ie\ very much in the center of the mass region for which the $b\anti b$
mode $\hsm$ studies were performed.  For $\mt\sim 200\gev$, the
constant mass value is somewhat larger, in the vicinity of 130 GeV.
The behavior of the $\hl$ and $\hh$ squared couplings to
$b\anti b$ and $t \anti t$, relative to the $\hsm$, as a function of $\mha$
is illustrated in Fig.~\couplings. We see that at low $\mha$
the couplings of the $\hh$ are fairly SM-like, while after the crossing over
(at $\mha\sim 100-130\gev$)
it is the $\hl$ which has SM-like couplings.  Thus, whichever Higgs
has mass in the vicinity of $\mz$ (plus radiative corrections)
is roughly SM-like.

However, the $b\anti b$ coupling for the Higgs with mass near $\mz$
is not precisely the same as for the $\hsm$.  For $\mha\gsim 50\gev$
it is somewhat larger.  This has the useful consequence that the
$b\anti b$ branching ratios for the $\hh$ or $\hl$ can be nearer to
100\% than for a $\hsm$ of the same mass.  Space does not allow a detailed
exposition on the branching ratios of the $\hl$ and $\hh$.
These are well-known.
\REF\bratios{R. Bork, J.F. Gunion, H.E. Haber, A. Seiden, \prdj{46} (1992)
2040.
J.F. Gunion, H.E. Haber, and C. Kao,
\prdj{46} (1992) 2907.
V. Barger, M.S. Berger, A.L. Stange, and R.J.N. Phillips,
\prdj{45} (1992) 4128.}
\refmark{\everybody,\bratios}\
In general, $BR(\hl\rta b\anti b)$ is
very near 1 until $\mhsm$ becomes very close to
its maximum value (at large $\mha$) at which point $BR(\hl\rta b\anti b)$
declines somewhat as other modes enter the picture.
The behavior of $BR(\hh\rta b\anti b)$ as a function of $\mha$
is much more complicated and is highly dependent upon $\tanb$.
In particular, $\hh\rta \hl\hl$ severely suppresses
the $\hh\rta b\anti b$ branching ratio for much of the relevant
$\mha$ range when $\tanb$ is $\sim 2$.
For larger $\tanb$, there is a window of moderate $\mha$ for which
$BR(\hh\rta b\anti b)$ is large.
To illustrate, in Fig.~\brbbbar\ we plot $BR(\h\rta b\anti b)$
as a function of $\mh$ ($\h=\hl$ or $\hh$) for $\mt=150\gev$, and
$\tanb=2$ and 5.

\smallskip
\noindent{\bf 2. Results and Discussion}
\smallskip

\REF\erice{J.F. Gunion, preprint UCD-93-8 (1993),
to appear in the Proceedings of {\it Properties of
SUSY Particles}, INFN workshop, eds. L. Cifarelli and A. Zichichi,
Erice, Sicily, October (1992).}

\FIG\surveyhbbi{Discovery contours (at the $4\sigma$ level) in $\mha$--$\tanb$
parameter space for the SSC with $L=30\fbi$ and LEP-200 with
$L=500\pbi$ for the reactions: a) $\epem\rta
\hl Z$ at LEP-200;
e) $W\hl X\rta l\gam\gam X$;
g) $t\rta \hp b$; h) $t\anti t \hh$, with $\hh\rta b\anti b$;
and i) $t\anti t \hl$, with $\hl\rta b\anti b$.
The contour corresponding to a given reaction is labelled by
the letter assigned to the reaction above. In each case, the letter
appears on the side of the contour for which detection of the
particular reaction {\it is} possible.
The letter assignments are chosen to be consistent with those in
J. Gunion and L. Orr, Ref.~\everybody, and Refs.~\pergunion\ and \erice,
in order to facilitate comparison with earlier results.
The large $\times$ indicates the location of the window
where no MSSM Higgs could be discovered at LEP-II or the SSC
without processes h) and i).
We have taken $\mt=150\gev$, $\mstop=1\tev$ and neglected squark mixing.
Scenario (A) refers to the notation established in Ref.~\erice; it corresponds
to the case in which charginos and neutralinos are taken to be heavy.}
\midinsert
\vbox{\phantom{0}\vskip 8in
\phantom{0}
\vskip .5in
\hskip -97pt
\special{ insert user$1:[jfgucd.rcsusyhiggs]hbb_susy_fig3.ps}
\vskip -1.95in }
{\rightskip=3pc
 \leftskip=3pc
 \Tenpoint\baselineskip=12pt%\parindent=1pc
\noindent Figure~\surveyhbbi:
Discovery contours (at the $4\sigma$ level) in $\mha$--$\tanb$
parameter space for the SSC with $L=30\fbi$ and LEP-200 with
$L=500\pbi$ for the reactions: a) $\epem\rta
\hl Z$ at LEP-200;
e) $W\hl X\rta l\gam\gam X$;
g) $t\rta \hp b$; h) $t\anti t \hh$, with $\hh\rta b\anti b$;
and i) $t\anti t \hl$, with $\hl\rta b\anti b$.
The contour corresponding to a given reaction is labelled by
the letter assigned to the reaction above. In each case, the letter
appears on the side of the contour for which detection of the
particular reaction {\it is} possible.
The letter assignments are chosen to be consistent with those in
J. Gunion and L. Orr, Ref.~\everybody, and Refs.~\pergunion\ and \erice,
in order to facilitate comparison with earlier results.
The large $\times$ indicates the location of the window
where no MSSM Higgs could be discovered at LEP-II or the SSC
without processes h) and i).
We have taken $\mt=150\gev$, $\mstop=1\tev$ and neglected squark mixing.
Scenario (A) refers to the notation established in Ref.~\erice; it corresponds
to the case in which charginos and neutralinos are taken to be heavy.}

\endinsert

\FIG\surveyhbbii{Discovery contours (at the $4\sigma$ level) in $\mha$--$\tanb$
parameter space for the SSC with $L=30\fbi$ and LEP-200 with
$L=500\pbi$ for the reactions: a) $\epem\rta
\hl Z$ at LEP-200; b) $\epem\rta \hl\ha$ at LEP-200; c) $\hl\rta 4l$;
e) $W\hl X\rta l\gam\gam X$;
h) $t\anti t \hh$, with $\hh\rta b\anti b$;
and i) $t\anti t \hl$, with $\hl\rta b\anti b$. Conventions
as in Fig.~\surveyhbbi, except that $\mt=200\gev$.
}
\midinsert
\vbox{\phantom{0}\vskip 8in
\phantom{0}
\vskip .5in
\hskip -97pt
\special{ insert user$1:[jfgucd.rcsusyhiggs]hbb_susy_fig4.ps}
\vskip -1.95in }
{\rightskip=3pc
 \leftskip=3pc
 \Tenpoint\baselineskip=12pt%\parindent=1pc
\noindent Figure~\surveyhbbii:
Discovery contours (at the $4\sigma$ level) in $\mha$--$\tanb$
parameter space for the SSC with $L=30\fbi$ and LEP-200 with
$L=500\pbi$ for the reactions: a) $\epem\rta
\hl Z$ at LEP-200; b) $\epem\rta \hl\ha$ at LEP-200; c) $\hl\rta 4l$;
e) $W\hl X\rta l\gam\gam X$;
h) $t\anti t \hh$, with $\hh\rta b\anti b$;
and i) $t\anti t \hl$, with $\hl\rta b\anti b$. Conventions
as in Fig.~\surveyhbbi, except that $\mt=200\gev$.
}

\endinsert

Results for the SSC are displayed in Figs.~\surveyhbbi\ and \surveyhbbii,
for $\mt=150$ and $200\gev$, respectively.
These figures display the regions of $\mha$--$\tanb$ parameter
space where detection of $t\anti t\h$ with $\h\rta b\anti b$ is possible
at the $4\sigma$ level for an integrated luminosity of $L=30\fbi$
--- the region of viability for $\h=\hl$ is indicated by the letter i),
while that for $\h=\hh$ is indicated by h).
The viable regions for an assortment of several other Higgs signals
are also given in each case.  The power of the $t\anti t
b\anti b$ final state modes is immediately apparent.

In the case of $\mt=150\gev$, Fig.~\surveyhbbi, the $\hl$ (i) and
$\hh$ (h) $t\anti t b\anti b$ modes
are viable in nearly all of parameter space above $\mha\sim 50\gev$.
The only exception is the delicate cross-over point discussed in
the previous section of the paper, where the $\hh$ and $\hl$ switch roles.
For $\mt=150\gev$ the critical region is $\mha\sim 100-110\gev$.  For such
$\mha$ values neither the $\hh$ nor the $\hl$ has full SM-strength $t\anti t$
coupling (see Fig.~\couplings) and, strictly speaking, neither satisfies
the $4\sigma$ discovery criterion.  However, it should be noted that
for large $\tanb$ the $\hh$ and $\hl$ are nearly degenerate at the
switch-over.  Consequently, their mass peaks can be effectively lumped
together, and the gap between the $\hh$ and $\hl$ regions is
in reality not present for $\tanb\gsim10$.  The other curves given
in Fig.~\surveyhbbi\ serve primarily to define the window
(referred to in our introductory discussion), marked with a large $\times$,
for which detection of a MSSM Higgs boson at LEP-II or the SSC
was previously deemed to be quite difficult.
We see that the $t\anti t\hl$, $\hl\rta b\anti b$ mode is viable
throughout this window region.  In addition, we observe that if
the LEP-II $Z\hl$ detection mode is removed, all but a tiny sliver
of parameter space is covered by the $t\anti t b\anti b$ mode
of the $\hl$
or else by the $t\rta\hp b$ mode for $\hp$ detection.  In addition,
everywhere that $\hl$ detection in the $l\gam\gam X$ final states
(arising from $W\hl X$ final states in which $W\rta l\nu$
and $\hl\rta\gam\gam$) is possible, $\hl$ detection in the $t\anti t b\anti b$
final state is also possible.

At $\mt=200\gev$ (Fig.~\surveyhbbii),
the $t\anti t b\anti b$ mode does not cover as large
a fraction of parameter space. This is because the $\hl$ is substantially
heavier when $\mha$ and $\tanb$ are both large ($\mhl\gsim 130\gev$),
than in the $\mt=150\gev$ case.  Thus, $BR(\hl\rta b\anti b)$ declines
at large $\mha$ (see Fig.~\brbbbar) due to the onset of $WW^*$ and $ZZ^*$
channels, while the $t\anti t\hl$ cross section is also somewhat smaller.
However, there is no window of concern in this case; detection
of at least one MSSM Higgs boson is possible throughout all of
parameter space even without using the $t\anti t b\anti b$ modes.
In Fig.~\surveyhbbii, we have chosen to plot only those modes
that allow detection of the $\hl$ (in particular, $t\rta\hp b$
and $\hh\rta 4\ell$ modes are not displayed) in addition to the
$t\anti t b\anti b$ modes for the $\hl$ and $\hh$.  This figure
illustrates the substantial overlap among different
$\hl$ detection modes.  In particular, where modes c), e) and i) overlap,
we should be able to experimentally verify the relative strengths of
the $ZZ\hl$, $t\anti t\hl$ and $b\anti b\hl$ couplings.
In addition, we see that the $\hl$
can be discovered, either at LEP-II or the SSC for all but a window
of parameter space with $\tanb\gsim 3.5$ and $70\lsim\mha\lsim150\gev$.

For smaller values of $\mt$, \eg\ in the vicinity of 100 GeV,
the $t\anti t b\anti b$ detection mode for the $\hl$ remains viable over much
the same region of parameter space as illustrated
in Fig.~\surveyhbbi\ for $\mt=150\gev$.
This is because radiative corrections
to the $\hl$ mass are small, and the upper limit for $\mhl$ (reached for
$\mha\gsim 100\gev$)
lies between $\sim50\gev$ and $\mz$ for $\tanb$ between 2 and $\infty$.
Thus, despite the increase of $t\anti t$-related backgrounds
relative to those found for $\mt=150\gev$ {\it at any given Higgs mass},
$\mhl$ moves into a lower mass region where these backgrounds are actually
smaller and the statistical significances (for 100\% $b\anti b$ branching
ratio)
quoted earlier are larger.
The region of viability for the $\hh$, whose mass remains in
the vicinity of $\mz$ for $\mha\lsim 100\gev$, does decrease somewhat
at $\mt=100\gev$ relative to $\mt=150\gev$
because of the increased $t\anti t$-related background levels.
Further details will be given in a longer paper.
\Ref\longer{J. Dai, J.F. Gunion and R. Vega, work in progress.}
Of course, current indications from both the Tevatron
and LEP are that $\mt$ is very likely to be $\gsim130-140\gev$.
If $\mt$ were to turn out to be near 100 GeV, the $\hl$
mass is sufficiently small that LEP-II will have an excellent chance
of detecting it in any case.

For $L=100\fbi$, the LHC is roughly equivalent to the SSC at $L=30\fbi$,
assuming the same efficiency and purity of $b$-tagging. However,
the efficiency of $b$-tagging at the LHC, given the many overlapping
events expected, may not be as great as assumed here. Other experimental
problems could arise regarding overlapping jets.

In the above, we have ignored the CP-odd $\ha$.  This is because
for both $\mt=150$ and $200\gev$
(and for $\tanb\gsim 0.5$) detection of the $\ha$ in the $t\anti t b\anti b$
mode is confined to a small area with $\tanb\lsim 2$ and $\mha$
below roughly 130 GeV. The upper limit is given by the
onset of the $\ha\rta Z\hl$ decay mode. Exactly how
low we can go in $\mha$ at low $\tanb$ values is not currently known.
We estimate that the results of Ref.~\previous\ can be safely extrapolated
down to 50-60 GeV.  But extension below that would require more detailed
study.  The inability to detect the $\ha$ for $\tanb\gsim 2$
is simply a consequence of the fact that the $\ha t\anti t$ coupling
is suppressed by $1/\tanb$. Thus, by $\tanb=2$ the $\ha t\anti t$
cross section is at most 1/4 of SM size.\foot{Actually, the $\gamma_5$
coupling also makes the cross section shape different as a function
of Higgs mass, so that this scaling argument is not precise.}

Of course, in this letter we have focused on the case where supersymmetric
partner masses are assumed to be large.  The results for the $t\anti t b\anti
b$ discovery modes of the $\hl$ and $\hh$ are essentially unmodified
unless the chargino and/or neutralino masses lie below 50-60 GeV.
This is because the $t\anti t b\anti b$ mode is mainly viable for
regions of parameter space such that the Higgs ($\hl$ or $\hh$)
mass is of order 80-130 GeV.  If the squark mass is also taken to be light,
then the region of viability for the $t\anti t b\anti b$ mode
at $\mt=200\gev$ actually expands since the $\hl$ has mass $\lsim\mz$,
and thus will have nearly 100\% branching ratio to $b\anti b$, as well
as being in the mass region of maximal sensitivity for the $t\anti t
b\anti b$ mode. Of course, other SM final state detection modes do suffer
if neutralinos and/or charginos are light.
As detailed in Ref.~\erice, the $\hh\rta4\ell$ mode is especially vulnerable.
The power of the $t\anti t b\anti b$ detection modes, however, is such
that the SSC alone remains able to see at least one MSSM Higgs boson
throughout essentially all of MSSM parameter space.

\smallskip
\noindent{\bf 3. Conclusion}
\smallskip

We have demonstrated that $b$-tagging can be used to isolate
$t\anti t \h$ events, in which $\h\rta b\anti b$, for $\h=\hl$ or
$\hh$ over a substantial range of the parameter space of the Minimal
Supersymmetric Model.  With inclusion of this detection mode,
for an accumulated luminosity of $L=30\fbi$
the SSC alone is guaranteed to find at least one (and more probably
several) of the MSSM Higgs bosons, for any value of $\mt\gsim 140\gev$.

\smallskip\noindent{\bf 4. Acknowledgements}
\smallskip
This work has been supported in part by Department of Energy
grants \#DE-FG03-91ER40674 and \#DE-AC03-76SF00515,
by Texas National Research Laboratory grants \#RGFY93-330
and \#RCFY93-229, and by National Science Foundation grant
NSF-PHY-88-18535.
JFG would like to thank L. Orr and H.E. Haber, with whom some
of the underlying programs employed for this project were developed.
\smallskip
\refout
\end